\documentclass[12pt,reqno]{article}
\usepackage{amsfonts}
\usepackage{amsmath}
\usepackage{amsbsy}
\usepackage{graphicx}
\usepackage{amssymb,latexsym}
 \numberwithin{equation}{section}

\begin{document}
 \allowdisplaybreaks[1]
\title{Effective Fluid FLRW Cosmologies of Minimal Massive Gravity}
\author{Nejat Tevfik Y$\i$lmaz\\
Department of Electrical and Electronics Engineering,\\
Ya\c{s}ar University,\\
Sel\c{c}uk Ya\c{s}ar Kamp\"{u}s\"{u}\\
\"{U}niversite Caddesi, No:35-37,\\
A\u{g}a\c{c}l\i Yol, 35100,\\
Bornova, \.{I}zmir, Turkey.\\
\texttt{nejat.yilmaz@yasar.edu.tr}} \maketitle
\begin{abstract}
By using a solution ansatz we partially decouple the metric and
the St\"{u}ckelberg sectors of the minimal massive gravity (MMGR).
In this scheme for a diagonal physical metric we find the general
solutions for the scalars of the theory and the particular
fiducial (background) metric which leads to these solutions. Then
we adopt this general formalism to construct the derivation of new
Friedmann-Lemaitre-Robertson-Walker (FLRW) cosmologies of the
theory in the presence of a so-called effective ideal fluid which
arises from our solution ansatz as a modifying, non-physical
source for the Einstein and the corresponding Friedmann equations.
\end{abstract}

\section{Introduction}
The de Rham, Gabadadze, Tolley (dRGT) massive gravity
\cite{dgrt1,dgrt2} which is Boulware-Deser (BD) \cite{BD1,BD2}
ghost-free and a non-linear continuation of the Fierz-Pauli
\cite{fp} massive gravity theory is reformulated in a series of
works \cite{hr1,hr2,hr3} to accommodate a general background or
fiducial metric in a minimal or a more general formalism. These
theories which include the dRGT theory as a special case are also
proven to be ghost-free at all orders. Both for the minimal and
the general cases cosmological solutions of these theories have
been intensively studied in the recent literature. The reader may
refer to a fair review of these works as well as their
achievements and shortcomings in \cite{derham}. In particular, it
has been shown that for flat background metric choice in spite of
the existence of spatially open FLRW solutions \cite{gum,gum2}
there exists no spatially flat or closed FLRW solutions \cite{mc}.
On the other hand there exit homogeneous and isotropic solutions
for the de Sitter \cite{ln} and the FLRW type \cite{higuchi}
background or fiducial metrics.

In the following work we focus on the minimal massive gravity
\cite{hr1}. We propose an ansatz which contributes an arbitrary
effective energy-momentum tensor to the metric sector which admits
Einstein equations modified by the existence of this effective
non-physical matter originated source. In this manner the metric
sector contains neither the St\"{u}ckelberg scalars nor the
fiducial metric explicitly. We later prove that when one considers
a general diagonal physical metric solution of the corresponding
Einstein equations the St\"{u}ckelberg sector can be exactly
solved for a functionally parametrized set of background metrics.
In doing this unlike the common approach of pre-determining the
background metric in the above mentioned literature we deduce it
from the solution ansatz as a function of the physical metric
solved via the Einstein equations and the arbitrarily chosen
effective energy-momentum source (which is arbitrary up to a
continuity equation). We determine the ansatz satisfying fiducial
metrics up to a family of integrable functions which lead also to
the solutions of the St\"{u}ckelberg scalars. Within this
formalism as a physical example we present the solutions of the
background metric and the St\"{u}ckelberg scalars as a function of
the scale factor and the arbitrarily fixed effective
energy-momentum tensor for the homogeneous and the isotropic FLRW
physical metric. We also derive the corresponding Friedmann and
the acceleration equations which only differ from the
GR-originating ones by the presence of an effective ideal fluid
arising from the solution ansatz we proposed which also can be
considered as a form of gravitational matter rather than a
physical one.

In Section two we discuss the details of our solution mechanism.
Section three contains the derivation of the solutions of the
St\"{u}ckelberg scalars and the background fiducial metric for a
diagonal choice of the physical metric. Finally, in Section four
we adopt the results of Section three to construct the FLRW
cosmological solutions of the minimal massive gravity within the
formalism introduced in Section two.
\section{The set-up}
 The minimal ghost-free massive gravity action coupled to matter via \cite{hr1}
 is
\begin{equation}\label{e1}
 S_{MMGR+MATT}=-M_p^2\int\bigg[ R\ast 1+2m^2tr(\sqrt{\Sigma})\ast 1+\Lambda^{\prime}\ast
 1\bigg]+S_{MATT},
\end{equation}
 with $M_{p}$ being the planck mass, $m$ is the graviton mass, $R$ is the Ricci scalar, and the $\ast$ is the Hodge star
 operator. In the above action $\Lambda^{\prime}=\Lambda-6m^2$. The four by four matrix $\Sigma$
 is
\begin{equation}\label{e2}
(\Sigma)^{\mu}_{\nu}=g^{\mu\rho}\partial_{\rho}\phi^a\partial_{\nu}\phi^{b}\bar{f}_{ab}(\phi^{c}).
\end{equation}
Here $g^{\mu\nu}$ is the inverse metric. $\phi^{a}$ are the
St\"{u}ckelberg scalar fields and $\bar{f}_{ab}(\phi^{c})$ is the
fiducial background metric yet not specified and which can
arbitrarily be chosen in \eqref{e1} in a relative physical context
to generate appropriate physical solutions. The spacetime indices
$\mu,\nu\cdots$ as well as the St\"{u}ckelberg indices
$a,b,c\cdots$ run on $\texttt{0,1,2,3}$. We have also introduced
the square root matrix obeying $\sqrt{\Sigma}\sqrt{\Sigma}=\Sigma$

Now, referring to \cite{hr1} we can write the metric equation
corresponding to \eqref{e1} as
\begin{equation}\label{e3}
R_{\mu\nu}-\frac{1}{2}g_{\mu\nu}R-\frac{1}{2}\Lambda
g_{\mu\nu}+\frac{1}{2}m^2\big[g\sqrt{\Sigma}+\sqrt{\Sigma}^{\:T}g\big]_{\mu\nu}+m^2g_{\mu\nu}(3-tr\big[\sqrt{\Sigma}\:\big])=G_NT_{\mu\nu},
\end{equation}
where $T_{\mu\nu}$ is the physical energy-momentum contribution of
the matter term in \eqref{e1}. The St\"{u}ckelberg scalar field
equations can also equivalently be achieved from the covariant
derivative-constancy of the metric equation as in \cite{hr1}, they
read
\begin{equation}\label{e4}
\nabla_{\mu}\big(\big[\sqrt{\Sigma}\:\big]^{\mu}_{\nu}+\big[g^{-1}\sqrt{\Sigma}^{\:T}g\big]^{\mu}_{\nu}-2tr\big[\sqrt{\Sigma}\:\big]\delta^{\mu}_{\nu}\big)
=0,
\end{equation}
here $\nabla_{\mu}$ is the covariant derivative corresponding to
the Levi-Civita connection of the metric $g$. It acts on
$\sqrt{\Sigma}$ as a (1,1) tensor. Following an extended but a
similar scheme like in \cite{constmmgr}  we will decouple the
Einsteinian gravity sector from the massive one by introducing a
solution ansatz of the form
\begin{equation}\label{e5}
\frac{1}{2}\big[g\sqrt{\Sigma}+\sqrt{\Sigma}^{\:T}g\big]-tr\big[\sqrt{\Sigma}\:\big]g=C_1g+C_2\tilde{T},
\end{equation}
where $C_1$ is a dimensionless, and $ C_2$ is a dimensionful
arbitrary constants which can be used to physically tune our
solution model to various forms. \eqref{e5} is written as a matrix
relation and $\tilde{T}$ is a four by four arbitrary (at this
stage) matrix functional which will play the role of effective
matter in the Einstein or the metric sector which we will call
dynamical. To appreciate our terminology of dynamical-kinematical
decoupling between the physical metric $g$ and the St\"{u}ckelberg
scalars and the fiducial metric it is enough to substitute
\eqref{e5} in \eqref{e3} which yields
\begin{equation}\label{e6}
R_{\mu\nu}-\frac{1}{2}g_{\mu\nu}R-\tilde{\Lambda}
g_{\mu\nu}=G_NT_{\mu\nu}-C_2m^2\tilde{T}_{\mu\nu},
\end{equation}
where we define
$\tilde{T}_{\mu\nu}\equiv[\tilde{T}]^{\mu}_{\:\:\:\nu}$. We have
the effective cosmological constant
\begin{equation}\label{e6.1}
\tilde{\Lambda}=\frac{1}{2}\Lambda-3m^2-C_1m^2.
\end{equation}
Now, the effective contribution of the ansatz \eqref{e5} to the
metric sector is more explicit in \eqref{e6} as these equations
are the usual Einstein equations for the metric $g$ which have a
yet arbitrary effective energy-momentum tensor addition on the
right hand side. We have called $\tilde{T}$ an energy-momentum
tensor because if we want the solutions of our ansatz \eqref{e5}
to be also the solutions of the scalar field equations \eqref{e4}
then $\tilde{T}$ must satisfy the constraint
\begin{equation}\label{e7}
\nabla^{\mu}\tilde{T}_{\mu\nu}=0.
\end{equation}
This can easily be seen from \eqref{e5} by applying a covariant
derivative on both sides
\begin{equation}\label{e8}
\nabla^{\mu}\big[\frac{1}{2}\big[g\sqrt{\Sigma}+\sqrt{\Sigma}^{\:T}g\big]-tr\big[\sqrt{\Sigma}\:\big]g\big]_{\mu\nu}=\nabla^{\mu}\big(
C_1g_{\mu\nu}+C_2\tilde{T}_{\mu\nu}\big),
\end{equation}
here if one uses the metric compatibility and imposes the
constraint \eqref{e7} on the right hand side then the left hand
side becomes zero. This result leads us to the scalar field
equations \eqref{e4} upon using the metric compatibility, and
index lowering. We have also used the fact that on functions
covariant derivative coincides with the ordinary one. Now, we will
focus on solving \eqref{e5}. If we multiply both sides in
\eqref{e5} by $2g^{-1}$ and then take the trace we find that
\begin{equation}\label{e9}
tr\big[\sqrt{\Sigma}\:\big]=-\frac{4}{3}C_1-\frac{1}{3}C_2\tilde{T}^{\mu}_{\:\:\:\mu},
\end{equation}
where we define
\begin{equation}\label{e10}
\tilde{T}^{\mu}_{\:\:\:\mu}\equiv
tr\big[g^{-1}\tilde{T}\:\big]=g^{\mu\rho}\tilde{T}_{\rho\mu}.
\end{equation}
When \eqref{e9} is substituted in \eqref{e5} one obtains
\begin{equation}\label{e11}
g\sqrt{\Sigma}+\sqrt{\Sigma}^{\:T}g=-\frac{2}{3}\big(C_1+C_2\tilde{T}^{\mu}_{\:\:\:\mu}\big)g+2C_2\tilde{T}.
\end{equation}
Furthermore, by using the symmetry of the matrix $g\sqrt{\Sigma}$
\cite{bac} we can write \eqref{e11} as
\begin{equation}\label{e11.5}
\sqrt{\Sigma}=-\frac{1}{3}\big(C_1+C_2\tilde{T}^{\mu}_{\:\:\:\mu}\big)\mathbf{1_{4}}+C_2g^{-1}\tilde{T}.
\end{equation}
Here $\mathbf{1_{4}}$ is the four dimensional unit matrix. If we
square both sides and isolate $f$ on the left hand side we
obtain
\begin{equation}\label{e12}
f=\mathcal{G}^{\prime},
\end{equation}
where we introduce $\mathcal{G}^{\prime}=g\mathcal{G}^2$ with
\begin{equation}\label{e13}
\mathcal{G}=-\big(\frac{C_1+C_2\tilde{T}^{\mu}_{\:\:\:\mu}}{3}\big)\mathbf{1_{4}}+C_2g^{-1}\tilde{T}.
\end{equation}
The constraint equation \eqref{e12} together with the effective
Einstein equations \eqref{e6} are the remainder equations of the
action \eqref{e1} to be solved upon specifying the effective
matter energy-momentum tensor $\tilde{T}$ subject to the
conservation constraint \eqref{e7}. We should remark that as we
have not specified the form of the physical metric $g$, and the
effective energy–momentum tensor $\tilde{T}$ \eqref{e12} is in its
most general matrix form which must be satisfied by
$f,g,\{\phi^a\},$ and $\tilde{T}$.
\section{Diagonal metric solutions}
In this section we will derive the general solutions of
\eqref{e12} for a diagonal choice of the physical and the induced
fiducial metrics. Firstly, let us state that when $g$ and $f$ are
assumed to be diagonal then the building blocks $\Sigma=g^{-1}f$
and $\sqrt{\Sigma}$ in \eqref{e1} also become diagonal. This
causes some degree of reduction of the non-linearity of the
gravity sector in \eqref{e1}. Next, if we focus on our solutions
we quickly observe via \eqref{e12} and \eqref{e13} that under the
diagonality assumption of the metrics the effective
energy-momentum tensor $\tilde{T}$ must also be chosen diagonal
for consistency. Now, in component form \eqref{e12} can be written
as
\begin{equation}\label{e14}
\partial_{\mu}\phi^a\partial_{\nu}\phi^{b}\bar{f}_{ab}(\phi^{c})=\mathcal{G}^{\prime}_{\mu\nu}.
\end{equation}
Let us also assume the fiducial metric diagonal too, namely
$\bar{f}=\text{diag}(f_{\texttt{00}},f_{ii})$ where $i=1,2,3$. In
this case the equations to be satisfied in \eqref{e14} take the
following form
\begin{subequations}\label{e15}
\begin{align}
&\sum\limits_{a=\texttt{0}}^{\texttt{3}}\big(\partial_{\mu}\phi^a)^2f_{aa}(\phi^b)=\mathcal{G}^{\prime}_{aa}\quad\forall\mu,\notag\\
\tag{\ref{e15}}\\
&\sum\limits_{a=\texttt{0}}^{\texttt{3}}\partial_{\mu}\phi^a\partial_{\nu}\phi^{a}f_{aa}(\phi^b)=0,\quad\text{when}\quad\mu\neq\nu.\notag
\end{align}
\end{subequations}
A straightforward solution to the second set of equations above
can be obtained by setting
\begin{equation}\label{e16}
\partial_{\mu}\phi^a\partial_{\nu}\phi^{a}=0,\quad\forall
a,\:\:\:\text{and}\:\:\:\mu\neq\nu.
\end{equation}
Now, as also discussed in \cite{constmmgr} one can multiply both
sides of the first set of equations in \eqref{e15} by
$\partial_{\alpha}\phi^{\alpha}\partial_{\beta}\phi^{\beta}\partial_{\gamma}\phi^{\gamma}$
(where there is no sum on the indices, and with
$\alpha,\beta,\gamma\neq\mu$ for all $\mu$) then successive use of
\eqref{e16} in the result leads to the set of equations
\begin{equation}\label{e17}
\big(\partial_{a}\phi^a)^2f_{aa}=\mathcal{G}^{\prime}_{aa}\quad\forall
a,
\end{equation}
where there is no sum on $a$ again. One can directly check that a
particular set of solutions to \eqref{e16} is obtained by setting
\begin{equation}\label{e18}
\partial_{\mu\neq a}\phi^{a}=0,\quad\forall \mu,a.
\end{equation}
Finally, now if we find a set of solutions for $\phi^a$'s and
$f_{aa}$'s which satisfy \eqref{e17} and \eqref{e18} for the
diagonal choice of $g,\bar{f},\tilde{T}$ then they also satisfy
the ansatz constraint equation \eqref{e12} so that they become
solutions of \eqref{e4} namely the St\"{u}ckelberg sector of
\eqref{e1}. Firstly, let us focus on Eq. \eqref{e18} which states
that for each $a=\texttt{0,1,2,3}$ the scalar field $\phi^a$ must
be a function of $x^a$ only. Thus, keeping this fact in mind if we
choose the diagonal fiducial metric components as
\begin{equation}\label{e20}
f_{aa}=\frac{\mathcal{G}^{\prime}_{aa}}{\big(F_a(x^a)\big)^2},
\end{equation}
where for each $a=\texttt{0,1,2,3}$ we have introduced the
arbitrary integrable functions $F_a$'s which only depend on the
corresponding coordinate $x^a$ then
\begin{equation}\label{e21}
\phi^{a}(x^{a}) =\pm \int F_{a}(x^{a})dx^{a},
\end{equation}
become the solutions of \eqref{e17}. One can instantly realize
that these scalar fields also satisfy \eqref{e18} thus,
\eqref{e16}. Therefore, they admit a family of solutions of the
constraint equation \eqref{e12} when the diagonal fiducial metric
components are chosen to be \eqref{e20}. On the other hand for a
diagonal choice of $g$ and $\tilde{T}$ we can explicitly construct
the diagonal matrix components $\mathcal{G}^{\prime}_{aa}$'s.
Firstly, let us observe
\begin{equation}\label{e22}
\tilde{T}^{\mu}_{\:\:\:\mu}=\sum\limits_{a=\texttt{0}}^{\texttt{3}}\frac{\tilde{T}_{aa}}{g_{aa}}.
\end{equation}
If we define
\begin{equation}\label{e23}
M=-\big(\frac{C_1+C_2\tilde{T}^{\mu}_{\:\:\:\mu}}{3}\big),
\end{equation}
then we have
\begin{equation}\label{e24}
\mathcal{G}^{\prime}_{aa}=M^2g_{aa}+2C_2M\tilde{T}_{aa}+C_2^2\frac{\tilde{T}_{aa}^2}{g_{aa}},
\end{equation}
where there is no sum on the index $a$ on the right hand side. We
should state once more that the off-diagonal elements of
$\mathcal{G}^{\prime}$ are zero. As a result we have shown that
for a diagonal metric solution of the modified Einstein equations
\eqref{e6} and a diagonal effective energy-momentum tensor source
introduced in the ansatz \eqref{e5} when \eqref{e20} is determined
via the substitution of \eqref{e24} then \eqref{e21} become the
solutions of the St\"{u}ckelberg sector of \eqref{e1}.
\section{Cosmological solutions}
In this section we will consider the FLRW metric ansatz for the
modified Einstein equations \eqref{e6}. Thus, as a special case of
the set of solutions constructed in the previous section we will
take the physical metric as the FLRW metric in spherical
coordinates
\begin{equation}\label{e25}
g=-dt^2+\frac{a^2(t)}{1-kr^2}dr^2+a^2(t)r^2 d\theta^2 +a^2(t)r^2
sin^2 \theta d\varphi^2,
\end{equation}
with $k$ being the scalar curvature of the $3-$space and $a(t)$
being the scale factor. Like the physical ideal fluid whose
energy-momentum tensor can be written as
$T=\text{diag}(\rho,p,p,p)$ in the momentarily co-moving frame,
for obtaining the homogeneous and isotropic metric solution
\eqref{e25} we also take the effective source of our ansatz as a
perfect fluid. Therefore, for the co-ordinate frame $\{x^{\mu}\}$
in which the metric components can be read via \eqref{e25} as
\begin{equation}\label{e25.1}
g_{\texttt{00}}=-1,\quad g_{\texttt{11}}=\frac{a^2}{1-kr^2},\quad
g_{\texttt{22}}=a^2r^2,\quad g_{\texttt{33}}=a^2r^2sin^2\theta,
\end{equation}
we can set the effective ideal fluid energy-momentum tensor from
its general definition as
\begin{equation}\label{e25.1}
\tilde{T}_{\mu\nu}=(\tilde{\rho}(t)+\tilde{p}(t))U_{\mu}U_{\nu}+\tilde{p}(t)g_{\mu\nu}.
\end{equation}
Thus, now if we take $U_{\texttt{0}}=1$, and $U_{\textit{i}}=0$
for $\textit{i}=1,2,3$ in the co-ordinate frame $\{x^{\mu}\}$ we
have
\begin{equation}\label{e26}
\tilde{T}=\text{diag}\big(\tilde{\rho}(t),\tilde{p}(t)g_{\texttt{11}},\tilde{p}(t)g_{\texttt{22}},\tilde{p}(t)g_{\texttt{33}}\big),
\end{equation}
with $\tilde{\rho}$ being the effective energy density, and
$\tilde{p}$ the effective pressure built out of the
St\"{u}ckelberg scalars which have an effective contribution to
the metric equation via the non-physical source \eqref{e26}.\\\\
\textsl{ The St\"{u}ckelberg sector}:\\\\
For the FLRW metric choice \eqref{e25} the contraction
$\tilde{T}^{\mu}_{\:\:\:\mu}=g^{\mu\nu}\tilde{T}_{\mu\nu}$ of
\eqref{e26} becomes
\begin{equation}\label{e27}
\tilde{T}^{\mu}_{\:\:\:\mu}=3\tilde{p}-\tilde{\rho}.
\end{equation}
The components of $\mathcal{G}^{\prime}$ which enter into the
definition of the fiducial metric \eqref{e20} that give rise to
the St\"{u}ckelberg field solutions \eqref{e21} can be explicitly
given as
\begin{subequations}\label{e28}
\begin{align}
\mathcal{G}^{\prime}_{\texttt{00}}&=-M^2+2C_2M\tilde{\rho}-C_2^2\tilde{\rho}^2,\notag\\\tag{\ref{e28}}\\
\mathcal{G}^{\prime}_{\textit{ii}}&=(M^2+2C_2M\tilde{p}+C_2^2\tilde{p}^2)g_{\textit{ii}}
\notag,
\end{align}
\end{subequations}
where $\textit{i}=1,2,3$ again. When the nature of the effective
ideal fluid composing $\tilde{T}$ is arbitrarily and independently
specified in addition to the physical ideal fluid, the scale
curvature $k$ is fixed and the scale factor $a(t)$ is solved from
the modified Einstein equations \eqref{e6} for the metric ansatz
\eqref{e25} these functions explicitly determine the fiducial
metric \eqref{e20} for which the scalar fields \eqref{e21} form up
the solutions of the St\"{u}ckelberg sector together with the
metric solution \eqref{e25} of the metric sector which we have
dynamically decoupled from the scalars.\\\\
\textsl{ The FLRW Dynamics}:\\\\
Now, if we use the FLRW metric \eqref{e25} in the on-shell metric
equation \eqref{e6} which is in Einstein form up to modification
by an effective source of energy-momentum tensor which in our case
we take to be a perfect fluid like the physical matter then we get
the fist Friedmann equation
\begin{equation}\label{e28.5}
H^2\equiv\big(\frac{\dot{a}}{a}\big)^2=\frac{G_N}{3}\rho-\frac{C_2m^2}{3}\tilde{\rho}-\frac{k}{a^2}-\frac{\tilde{\Lambda}}{3},
\end{equation}
and via the first and the second Friedmann equations the
acceleration equation
\begin{equation}\label{e29}
\frac{\ddot{a}}{a}=-\bigg[\frac{G_N(\rho+3p)-C_2m^2(\tilde{\rho}+3\tilde{p})}{6}\bigg]-\frac{\tilde{\Lambda}}{3}.
\end{equation}
When the cosmological metric ansatz \eqref{e25} is used in the
 physical perfect fluid matter energy-momentum conservation equation
 $\nabla^{\mu}T_{\mu\nu}=0$, and also in the similar constraint equation
 \eqref{e7} for the effective perfect fluid
 \eqref{e26} we obtain the continuity or the fluid equations
\begin{equation}\label{e30}
\dot{\rho}+3H(\rho+p)=0\quad,\quad\dot{\tilde{\rho}}+3H(\tilde{\rho}+\tilde{p})=0,
\end{equation}
respectively. Therefore, when the equation of states are
determined for the physical and the effective ideal fluids from
\eqref{e28.5}, \eqref{e29}, \eqref{e30} one can solve the scale
factor $a(t)$ and the energy densities $\rho(t)$ and
$\tilde{\rho}(t)$. Then one can use these solutions in \eqref{e27}
and \eqref{e28} to explicitly obtain the building blocks of the
fiducial metric \eqref{e20}. After specifying the arbitrary
integrable functions $\{F_a(x^a)\}$ one entirely determines the
background metric, and finds the St\"{u}ckelberg scalar solutions
from \eqref{e21}. As a result one obtains the cosmological FLRW
metric, and the St\"{u}ckelberg scalar field solutions of
\eqref{e1} for the specially constructed fiducial metric. We have
to stress that these solutions work for any scale curvature
without putting restrictions on it. Before concluding we have to
make an important remark regarding our solution generation method;
the reader should observe that although the equation of state of
the matter ideal fluid is determined by its physical nature the
effective ideal fluid which emerges as a mathematical source of
our solution ansatz \eqref{e5} has no restrictions on it apart
from \eqref{e7} which leads to the second equation in \eqref{e30}.
For example the physical perfect fluids have the equation of state
in the form $p=w\rho$, however, our construction suggests no
special form for the equation of state for the effective ideal
fluid, namely it seems that one can assign any form to it
$\tilde{p}=\textit{f}(\tilde{\rho})$.
\section{Conclusion}
After decoupling completely the metric sector of the minimal
massive gravity from the St\"{u}ckelberg scalar fields of the mass
term by introducing an ansatz we have focussed on the solutions
satisfying the ansatz constraint on the background metric and the
St\"{u}ckelberg fields. We have shown that in this on-shell
formalism the metric field equations reduce to the Einstein
equations modified solely by a contribution of an effective
energy-momentum tensor which appears as a source in the ansatz we
have proposed. We have later constructed a formal family of
solutions for the background fiducial metric and the
St\"{u}ckelberg fields in the presence of a diagonal physical
metric. In the last section we have considered the cosmological
applications within this scheme. By assuming the homogeneous and
isotropic FLRW metric in the modified Einstein equations where the
scalars do not appear explicitly we have given the corresponding
solutions of the scalars and the background metric which satisfy
the ansatz constraint and lead to the above-mentioned decoupling.
Due to the necessity of homogeneity and the isotropy of the FLRW
cosmology we have restricted ourselves to an ideal fluid form for
the effective matter. Then via the FLRW metric used in the
modified Einstein sector we have also derived the cosmological
equations for the scale factor dynamics, namely the first
Friedmann and the acceleration equations.

The method of solving the minimal massive gravity field equations
that is constructed in this work diverges from the one widely used
in the literature. Instead of pre-determining the background
metric from the start we find it as a solution of an ansatz which
enables the decoupling of the scalars from the metric equation.
Thus, we design the necessary fiducial metric to generate
solutions to the theory in this sense. We have performed this
analysis first for a more general diagonal physical metric then as
a special case for the FLRW cosmological one. In this manner,
after determining the background fiducial metric we also obtain
the scalar field solutions when the physical metric is specified.
We should state that apart from constructing explicit solutions
this kind of formalism may have essential physical consequences.
The effective matter energy-momentum tensor which appears as an
overall mathematical source within the solution constraint is
entering in the cosmological equations as completely an arbitrary
effective ideal fluid whose equation of state can be chosen in any
form to generate various solutions. This may be interesting in two
aspects; firstly, one can design the necessary forms of effective
matter (design the corresponding solution instead) to cure the
dark energy or the dark matter problems in the presence of
accompanying physical matter, secondly, one can work out various
exotic solutions arising from this new kind of gravitational
matter. If on the other hand, one considers solely the
self-acceleration issue we observe that although in our solution
scheme the modified on-shell metric equation has an effective
cosmological constant the new coming terms are suppressed by the
graviton mass. However, it is also obvious that having an
effective ideal fluid source degree of freedom in the cosmological
equations guaranties the existence of self-accelerating solutions
in the absence of the cosmological constant. The solution
construction formalism proposed in this work can also be extended
to cover the more general massive gravity case whose Lagrangian
has more non-linearly involved mass terms. In this manner, one can
also derive, and study new cosmological solutions of the general
massive gravity. We leave this discussion here to appear in a
separate work \cite{gencosmo}. However, in this direction we must
also stress on the problems arising for the known types of
cosmological solutions of massive gravity \cite{gum3,gum4}, as
well as the problems that are reported via the inspection of the
theory from a more general perspective \cite{des}.

\end{document}